\documentclass[twocolumn,floatfix]{revtex4}
\pdfoutput=1
\usepackage{graphicx}
\usepackage{dcolumn}
\usepackage{longtable}
\usepackage{amsmath}
\usepackage{amssymb}
\usepackage{color}
\usepackage{pdflscape}

\begin{document}

\title
{Resolution of the spin paradox in the Nilsson model}

\author
{Hadi Sobhani$^1$}
\email[Corresponding author: ]{hadisobhani8637@gmail.com}

\author
{Hassan Hassanabadi$^1$}

\author
{Dennis Bonatsos$^2$}

\affiliation
{$^1$Faculty of Physics, Shahrood University of Technology, Shahrood, Iran, P.O. Box 3619995161-316}

\affiliation
{$^2$Institute of Nuclear and Particle Physics, National Centre for Scientific Research 
``Demokritos'', GR-15310 Aghia Paraskevi, Attiki, Greece}

\begin{abstract}

In the well-known Nilsson diagrams, depicting the dependence of the nuclear single-particle energy levels on quadrupole deformation, a spin paradox appears {as the deformation sets in, leading from spherical shapes to prolate deformed shapes with cylindrical symmetry}. 
Bunches of levels corresponding to a spherical shell model orbital, sharing the same orbital angular momentum and the same total angular momentum, appear to correspond to Nilsson energy levels, labeled by asymptotic quantum numbers in cylindrical coordinates, some of which have spin up, while some others have spin down. Furthermore, for some orbitals the correspondence between spherical shell model quantum numbers and Nilsson asymptotic quantum numbers is not the same for protons and for neutrons. Introducing a new rule of correspondence between the two sets of quantum numbers, we show that the spin paradox is resolved and full agreement between the proton and neutron Nilsson diagrams is established. The form of the Nilsson diagrams as a function of the quadrupole deformation remains unchanged, the only difference between the new diagrams and the traditional ones being the mutual exchange of the Nilsson labels for certain pairs of single-particle energy levels.     

\end{abstract}

\maketitle

\section{Introduction}

The nuclear shell model \cite{Mayer,MJ,Brussaard,Heyde,Talmi} is an appropriate tool for reproducing the observed magic numbers in atomic nuclei, which correspond to unusual stability of nuclei possessing these particular numbers of protons and/or  neutrons. This model is based on the assumption that each nucleon is moving inside the nucleus in an average potential due to the other nucleons, resembling a three-dimensional (3D) isotropic harmonic oscillator (HO) potential with a flattened bottom, to which the spin-orbit interaction is added, playing a crucial role in changing the 3D HO \cite{Wybourne,Smirnov,IacLie} magic numbers, 2, 8, 20, 40, 70, 112, \dots into the nuclear magic numbers, 2, 8, 20, 28, 50, 82, 126, \dots,  observed experimentally. The single-particle energy levels, {i.e., the nucleon energy levels within the 3D isotropic HO potential,} are filled in accordance to the Pauli  exclusion principle. Spherical coordinates are used for the description of the atomic nucleus. Each energy level characterized by total angular momentum $j$ can accommodate up to $ 2(2j+1) $ protons or neutrons. Groups of levels forming a magic number are supposed to be inert, while the nuclear properties are decided by the protons and neutrons in incomplete shells, called the valence protons and neutrons. {The picture described so far corresponds to nuclei with few valence protons and neutrons, therefore lying near to closed shells. These nuclei are known to have spherical or near-spherical shapes \cite{BM1}}. 
 
{Moving away from closed shells, the nucleus acquires several valence protons and neutrons, and quadrupole deformation sets in \cite{BM}. As a consequence,}  the order of the single-particle energy levels is modified. The levels may cross each other. For the first time, Nilsson \cite{Nilsson,MoNi,MotNil,NilssonP4,NilssonSH,Ragnarsson,RN,NR,Mottelson,Casten} formulated such a study, 
using a modified 3D anisotropic HO with axially symmetric quadrupole deformation \cite{Takahashi,RD,ND,PVI,Lenis,Sugawara,ArimaJPG}, to which the spin-orbit interaction is added. Although the spherical coordinates remain convenient for small deformations, it has been realized that for large deformations the cylindrical coordinates become more appropriate, since they offer a suitable basis for the definition of asymptotic quantum numbers \cite{MN,Rassey,Quentin,Boisson}, which are exact for very large deformations but remain satisfactorily good even at moderate deformations. {The use of asymptotic quantum numbers for deformed nuclei is just an example of choosing a basis as close as possible to the dynamics of the physical system under study \cite{Welsh,Rowe}.}
It should be noted that the Nilsson model is still in wide use, 65 years after its introduction \cite{Nilsson}, either in the framework of Nilsson mean-field plus pairing calculations \cite{Guan1,Guan2}, cranking shell model calculations based on the Nilsson potential with pairing correlations \cite{Zhang}, or in relation to approximate SU(3) symmetries \cite{Kota} like the pseudo-SU(3) symmetry \cite{Adler,Shimizu,pseudo1,pseudo2,Ginocchio1,Ginocchio2,Quesne,Velazquez,DW1,DW2}, the quasi-SU(3) symmetry \cite{Zuker1,Zuker2}, the proxy-SU(3) symmetry \cite{proxy1,proxy2,proxy3,Sobhani,EPJASM,CRC,EPJAHW,EPJASC}, and the recovery of SU(3) symmetry at superdeformation \cite{Arima}. 

From the above it becomes clear that each single-particle energy level can be described by two sets of quantum numbers, an exact one in the spherical coordinates and an asymptotic one in the cylindrical coordinates. One can see this correspondence in the Nilsson diagrams \cite{Nilsson,NR,FS}, which are figures presenting the single-particle energies as a function of the quadrupole deformation. Looking into the details of this correspondence, one sees a {\sl spin paradox}: bunches of levels forming an orbital in the spherical basis, sharing the same orbital angular momentum $l$ and the same total angular momentum $j$,  correspond to a mixture of asymptotic levels in the cylindrical basis, some of them having spin up and some having spin down. Furthermore, the correspondence between spherical and cylindrical labels for the same single-particle level is not always the same for protons and for neutrons.     

In this paper, we are going to resolve the spin paradox by pointing out a different, well-defined way for making the correspondence between the spherical quantum numbers and the asymptotic cylindrical quantum numbers for each single-particle level. Within this new scheme,  bunches of levels forming an orbital in the spherical basis, sharing the same orbital angular momentum $l$ and the same total angular momentum $j$,  correspond to asymptotic levels in the cylindrical basis  having the same spin projection (up or down), and not a mixture of them. Furthermore, within this new scheme the correspondence between spherical and cylindrical labels for the same single-particle level is guaranteed to be the same for protons and for neutrons.

In Section 2 of the present work we review the description of the 3D isotropic HO in different coordinate systems, while in Section 3 a review of the Nilsson anisotropic, axially symmetric oscillator in the asymptotic cylindrical coordinates is given. The spin paradox is described in Section 4, while in Section 5 the new scheme leading to its resolution is introduced. Numerical results obtained within the new scheme are presented in Section 6, while in Section 7 the conclusions of this work are given.

\section{The three dimensional isotropic harmonic oscillator in different coordinate systems}

We start with a brief review of the description of the three-dimensional (3D) isotropic harmonic oscillator (HO) \cite{Wybourne,Smirnov,IacLie} in three different coordinate systems \cite{Greiner}, in order to introduce the notation, bearing in mind that the results for the measurable quantities of a physical system must be independent from the coordinate system used.

The potential of the isotropic HO is  written as \cite{Greiner}
\begin{multline}
\label{simple harmonic potential}
V(\mathbf{r}) = \frac{1}{2} m \omega^2 \left( x^2 + y^2 + z^2 \right) \\
= \frac{1}{2} m \omega^2 \left( \rho^2 + z^2 \right) 
= \frac{1}{2} m \omega^2 r^2,
\end{multline} 
where $ x, y, z $ are the Cartesian coordinates, while $ \rho = \sqrt{x^2 + y^2} $ and $ r = \sqrt{x^2 + y^2 + z^2} $ are the radial coordinates in the cylindrical and the spherical coordinate systems, respectively, and $\omega$ is the angular frequency of the oscillator. For the further details we refer to Refs. \cite{Greiner,GM}.
The energy eigenvalues in the three coordinate systems are given by  
\begin{align}
\label{E-cart}
E_{\text{Cart.}} =& \hbar \omega \left( n_x + n_y + n_z + \frac{3}{2} \right),\\
\label{E-cyl}
E_{\text{Cyl.}} =& \hbar \omega \left( 2 n_\rho + n_z + 
{|\Lambda|} 
+ \frac{3}{2} \right), \\
\label{E-sph}
E_{\text{Sph.}} =& \hbar \omega \left( 2 n_r + L + \frac{3}{2} \right),
\end{align}
where the quantum numbers in different directions are shown by $ n_i$, with $i=x,y,z,\rho,r$, while $L$ and $\Lambda$ represent the angular momentum and its projection on the $z$-axis  respectively. 

Separation of variables is possible in the Cartesian coordinates, therefore the wave function of the 3D isotropic HO  is simply the product of the three one-dimensional (1D) HO wave functions \cite{Greiner} 
\begin{multline}
\label{phi_x}
\phi_{n} (\xi) = N_n e^{-\frac{\xi^2}{2}} H_n (\xi), \\
{
N_n = \sqrt{\frac{1}{2^n n!} \sqrt{\frac{\lambda}{\pi}}}, \quad
\xi=x \sqrt{\lambda},
\quad
\lambda = \frac{m \omega}{\hbar} 
}
\end{multline}
where $H_n$ are the Hermite polynomials, reading 
\begin{align}
\psi_{n_x, n_y, n_z} = \phi_{n_x} (x) \phi_{n_y} (y)\phi_{n_z} (z).
\end{align}

The eigenfunctions of the harmonic oscillator in cylindrical coordinates are given by \cite{GM}
\begin{multline}
\psi_{n_z, n_\rho, \Lambda} = N_c \exp \left[-\frac{k}{2} \left( z^2 + \rho^2 \right)\right] \\ H_{n_z} (k z) \rho^{|\Lambda|} L_{n_{\rho}} ^{|\Lambda|} \left(k \rho^2\right) e^{i \Lambda \phi}
\end{multline}
where $L_n$ are the Laguerre polynomials, $N_c$ is a normalization constant  and $ k = m \omega / \hbar $. It is seen that the azimuthal part of this wave function is a general phase which does not affect the modules of the wave function. In other words, we have $ \left[ H, L_z \right] = 0 $,  which means $ \Lambda $ is a constant of the motion.

Finally, the wave function of the 3D isotropic HO in spherical coordinates is \cite{Greiner,GM} 
\begin{multline}
\psi_{n_r, L, \Lambda} = N_c r^L \exp \left[-\frac{m \omega}{2 \hbar}r^2\right] \\ _1F_1 \left( -n_r, L + \frac{3}{2}, \frac{m \omega}{\hbar} r^2 \right) Y_{L \Lambda} (\theta, \varphi),
\end{multline} 
where $ Y_{L \Lambda} (\theta, \varphi) $ are the spherical harmonics $ Y_{L \Lambda} (\theta, \varphi) $ and $ _1 F_1 $ denote the  confluent hypergeometric functions.
 
In principle, it is not important which system of coordinates is used for the description of nuclei, but depending on the available symmetries in the problem, the use of each system of coordinates has its own advantages. For instance, for small {$ (\delta \lesssim 0.1) $} and moderate {$ (0.1 \lesssim \delta \lesssim 0.3) $} deformations {(see Eq. (\ref{delt}) for the definition of the deformation $\delta$)} it is customary to use the spherical coordinates, because in these cases the deformations in the shape of a nucleus do not lead far from the spherical shape \cite{NR}. 
On the other hand, for large deformations {($\delta > 0.3$)}, where the nucleus assumes an axially symmetric shape, it is preferable to use cylindrical coordinates \cite{NR}. Nevertheless, the results must be independent from the selection of the coordinate system. {More details will be given in the next section.} 

\section{The modified harmonic oscillator}

The starting point of the shell model was the interpretation of the nuclear magic numbers, which are reproduced by adding  to the 3D isotropic HO the spin-orbit interaction, i.e., a term describing the coupling of the spin  with the orbital angular momentum \cite{Mayer,MJ}. Subsequently, the Nilsson model was introduced, in which the 3D isotropic HO is replaced by an axially symmetric 3D anisotropic HO with cylindrical symmetry. The Nilsson Hamiltonian reads \cite{Nilsson,NR}
\begin{equation}
H = H_{\text{osc}} + H',
\end{equation}
where $ H_{\text{osc}} $ is the Hamiltonian of the 3D anisotropic HO to be discussed below, while the correction terms are
\begin{equation}\label{Hprime}
H' = - 2 \kappa \hbar \omega_0  
\mathbf{L} \cdot \mathbf{s} - \mu \kappa \hbar \omega_0 \left( \mathbf{L}^2 - \left< \mathbf{L}^2 \right>_N  \right),
\end{equation}
with the first term representing the spin-orbit interaction and the second term, proportional to the square of the orbital angular momentum,  rounding up the bottom of the potenial. The term  
 $ \left< L^2 \right>_N  = N(N+3)/2 $ has a constant value in each shell, while $ \kappa $ and $ \mu $ are parameters that have different values within each proton or  neutron major shell, listed, for example,  in
Refs. \cite{NR,BM}.  

The Hamiltonian of the 3D anisotropic HO with cylindrical symmetry used in the Nilsson model is \cite{NR}
 \begin{equation}
H_{\text{osc}} = -\frac{\hbar^2}{2M} \Delta + \frac{M}{2} \left[ \omega_\perp ^2 \left(x^2+y^2\right) + \omega_z ^2 z^2
\right],
\end{equation}
where the frequencies in the $x$-$y$ plane and in the $z$-direction are denoted by $ \omega_\perp $ and $ \omega_z $, respectively, {$M$ is the nucleon mass, and $\Delta$ is the usual Laplacian operator}. Then the deformation parameter is defined in the form
\begin{equation}\label{delt}
\delta = \frac{\omega_\perp - \omega_z}{\omega_0}.
\end{equation}

The modified harmonic oscillator Hamiltonian can be solved in different ways, depending on the size of the deformation. For small deformations {($\delta \lesssim 0.1$)}, perturbation techniques within the spherical basis can be used \cite{NR}, while at large deformations {($\delta > 0.3$)} the asymptotic basis in cylindrical coordinates is used \cite{NR}.

{In the case of small deformations ($\delta \lesssim 0.1$), one can employ perturbation theory in the spherical coordinates, the result being \cite{NR} 
\begin{align}
\left< N L s j \Omega \left| \delta \hbar' \right| N L s j \Omega \right> = \frac{1}{6} \delta M \omega_0 ^2 \left< r^2 \right> \frac{3 \Omega^2 - j (j+1)}{j(j+1)}, 
\end{align}
where $ \hbar' $ is the first order of the perturbation, $N$ is the number of oscillator quanta, $L$ is the orbital angular momentum, $s$ is the spin, $j$ is the total angular momentum and $\Omega$ is its ptojection on the $z$-axis. }

According to our purpose in this paper, we start from the case of large deformations {($\delta> 0.3$)} and we show how the solutions obtained for large deformations can be used for obtaining solutions valid at small deformations {($\delta\lesssim 0.1$)}, providing in parallel the correct nuclear magic numbers, without making any change in the Hamiltonian, but by simply imposing a rule of correspondence between the cylindrical quantum numbers used for large deformations and the spherical quantum numbers used at small deformations.  
 
In the case of large deformations, it is convenient to introduce what one may call ``stretched'' coordinates \cite{NR}
\begin{equation}
\xi = x \sqrt{\frac{M \omega_\perp}{\hbar}},
\quad
\eta = y \sqrt{\frac{M \omega_\perp}{\hbar}},
\quad
\zeta = z \sqrt{\frac{M \omega_z}{\hbar}}.
\end{equation}
Consequently, the components of angular momentum are defined in terms  of the ``stretched'' coordinates as 
\begin{equation}
\left(L_t\right)_x = -i \hbar \left( \eta \frac{\partial}{\partial \zeta} - \zeta \frac{\partial}{\partial \eta}\right),
\end{equation}  {
\begin{equation}
\left(L_t\right)_y = -i \hbar \left( \zeta \frac{\partial}{\partial \xi} - \xi \frac{\partial}{\partial \zeta}\right),
\end{equation}
\begin{equation}
\left(L_t\right)_z = -i \hbar \left( \xi \frac{\partial}{\partial \eta} - \eta \frac{\partial}{\partial \xi}\right).
\end{equation}}
As a consequence, the correction terms in the Hamiltonian are taking the form
\begin{align}\label{Ham}
H' _{\text{def}} = - 2 \kappa \hbar \omega_0  
\mathbf{L}_t \cdot \mathbf{s} - \mu \kappa \hbar \omega_0 \left( \mathbf{L}_t ^2 - \left< L_t ^2 \right>_N  \right),
\end{align} 
with the index $ t $ dropped hereafter, according to the approximation introduced by Nilsson \cite{Nilsson,NR}. {In other words, Eq. (\ref{Ham}) is considered as the analog in the ``stretched'' coordinates of Eq. (\ref{Hprime}) in the usual cartesian coordinates. The parameters $\kappa$ and $\mu$ are the ones used in Eq. (\ref{Hprime}),   
while $\omega_0$ is slighlty dependent on the deformation, as seen from the relation obtained from the volume conservation condition \cite{Nilsson}
\begin{align}
	\omega_0 (\delta ) = 
	\overset{0}{\omega}_0 \left(
	1 - \frac{4}{3} \delta^2 - \frac{16}{27} \delta^3
	\right)^{-1/6},
\end{align}
where $ \overset{0}{\omega}_0 $ is $ \omega_0 (\delta = 0) $.}

In these coordinates the eigenfunctions and eigenvalues for $ H_{\text{osc}} $ are found to read 
\begin{multline}
\label{eigen function}
\psi_{n_z,n_\perp, \Lambda} (\rho, \varphi, z)= N_c e^{-\zeta^2 /2}H_{n_z} (\zeta)
\rho^{|\Lambda|} e^{-\rho^2 /2} \\ 
_1 F_1 \left( - \frac{n_\perp - |\Lambda|}{2}, |\Lambda|+1; \rho^2 \right) e^{i \Lambda \varphi}, \\
E_{n_z, n_\perp} = \hbar \omega_0 \left(
N + \frac{3}{2} + \left( n_\perp - 2n_z \right) \frac{\delta}{3} \right),  
\end{multline}
with $ \rho^2 = \xi^2 + \eta^2 $, $ N = n_\perp + n_z $, $ n_\perp = n_x + n_y $. When $H'_{\text{def}}$ is added, these functions are not  eigenfunctions of the total Hamiltonian $ H_{\text{tot}} = H_{\text{osc}} + H' _{\text{def}}  $. The mean value of $ \mathbf{L}^2 $ is taking a constant value within each shell, but for the other terms of $H'_{\text{def}}$ the matrix elements read \cite{Nilsson,NR}
\begin{multline}
\left< n_z-1, n_\perp+1, \Lambda+1, \Sigma-1 \left|
\mathbf{L} \cdot \mathbf{s}
\right|n_z, n_\perp, \Lambda, \Sigma
\right> =  \\ -\frac{1}{2} \sqrt{n_z \left(n_\perp + \Lambda + 2 \right)}, \\
\left< n_z+1, n_\perp-1, \Lambda+1, \Sigma-1 \left|
\mathbf{L} \cdot \mathbf{s}
\right|n_z, n_\perp, \Lambda, \Sigma
\right> = \\ \frac{1}{2} \sqrt{\left(n_z+1\right) \left(n_\perp - \Lambda\right)}, \\
\left< n_z+1, n_\perp-1, \Lambda-1, \Sigma+1 \left|
\mathbf{L} \cdot \mathbf{s}
\right|n_z, n_\perp, \Lambda, \Sigma
\right> = \\ -\frac{1}{2} \sqrt{\left(n_z+1\right) \left(n_\perp + \Lambda\right)}, \\
\left< n_z-1, n_\perp+1, \Lambda-1, \Sigma+1 \left|
\mathbf{L} \cdot \mathbf{s}
\right|n_z, n_\perp, \Lambda, \Sigma
\right> =\\ \frac{1}{2} \sqrt{n_z \left(n_\perp - \Lambda + 2 \right)},
\end{multline} 
and
\begin{multline}
\left< n_z, n_\perp, \Lambda, \Sigma \left|
\mathbf{L}^2 \right|n_z, n_\perp, \Lambda, \Sigma
\right> = \\2n_z \left( n_\perp + 1 \right) + n_\perp + \Lambda^2,\\
\left< n_z + 2, n_\perp - 2, \Lambda, \Sigma \left|
\mathbf{L}^2 \right|n_z, n_\perp, \Lambda, \Sigma
\right> =\\ - \sqrt{\left( n_z + 2 \right)  \left( n_z + 1 \right)  \left( n_\perp + \Lambda \right)  \left( n_\perp - \Lambda \right)},\\
\left< n_z - 2, n_\perp + 2, \Lambda, \Sigma \left|
\mathbf{L}^2 \right|n_z, n_\perp, \Lambda, \Sigma
\right> =\\ - \sqrt{\left( n_z -1 \right)   n_z \left( n_\perp + \Lambda + 2 \right)  \left( n_\perp - \Lambda + 2\right)},
\end{multline}
where $ \Sigma = \pm 1/2 $ is the projection of the spin and the notation $ \psi = \left|n_z, n_\perp, \Lambda, \Sigma \right> $ is used. The selection rules appearing in the above equations indicate that in order to obtain the eigenfunctions and eigenvalues of $H_{\text{tot}}$, the Hamiltonian matrix within each $ N $ shell should be constructed and subsequently diagonalized. 

Up to here, we have reviewed the asymptotic solutions of the Nilsson Hamiltonian, valid at large deformations {($\delta >0.3$)}. In Section 5, a rule will be presented, according to which each state is uniquely labeled both in the cylindrical and in the spherical coordinates, so that one can recover the nuclear magic numbers at zero deformation, as well as correct solutions for small {($\delta \lesssim 0.1$)} and moderate 
{($0.1\lesssim \delta \lesssim 0.3$)} deformations using the asymptotic quantum numbers $ \Omega,N, n_z, \Lambda, \Sigma $ with $ \Omega = \Lambda + \Sigma $. But before doing so, we are going to discuss the spin paradox appearing in the Nilsson model, which will be resolved by the just mentioned rule. 

\section{The spin paradox}

As we have already seen, Nilsson states in the asymptotic basis {\cite{Nilsson,NR}} are labeled as $\Omega[N n_z \Lambda]$, where $N$  is the total number of oscillator quanta and  $n_z$ is the number of quanta along the $z$-axis, while $\Lambda$ and $\Omega$ denote respectively the projections of the orbital angular momentum and the total angular momentum on the $z$-axis. {The full wave functions can be found on page 115 of Ref. \cite{NR}}.   On the other hand, in the spherical shell model basis orbitals are denoted by 
$n l j _\Omega$, where $l$ and $j$ are the quantum numbers associated respectively with the orbital angular momentum and the total angular momentum, and $n=n_r+1$. 

The correspondence between the asymptotic cylindrical Nilsson orbitals and the spherical shell model orbitals for neutrons, as taken from the standard Nilsson diagrams \cite{FS}, is shown in Table \ref{TA}. The following observations can be made, with the spherical orbitals involved in the discussion shown in boldface. 

a) In several cases the spherical shell model orbitals are made up from Nilsson orbitals of mixed spin projection $\Sigma$. For example, the 1g7/2 orbital is made out of three Nilsson orbitals with spin up (1/2[420], 
3/2[411], 5/2[402]) and one Nilsson orbital with spin down (7/2[404]). We call this fact {\sl the spin paradox}.

b) In the same cases as in a), the spherical shell model orbitals are made up from Nilsson orbitals of mixed $n_\rho$ values. For example, the 1g7/2 orbital is made out of three Nilsoon orbitals with $n_\rho=1$  (1/2[420], 
3/2[411], 5/2[402]) and one Nilsson orbital with $n_\rho=0$ (7/2[404]).

c) The cases of a) and b) are in on-to-one correspondence with cases of groups of Nilsson orbitals possessing mixed values of $n_r$. For example, in the group of Nilsson orbitals 1/2[431], 3/2[422], 5/2[413], 7/2[404], the first three correspond to $n_r=1$, while the last one to $n_r=0$.  

Furthermore, the correspondence between the asymptotic cylindrical Nilsson orbitals and the spherical shell model orbitals for protons, again taken from the standard Nilsson diagrams \cite{FS}, is shown in Table \ref{TB}. The following observations can be made.

a) The spin paradox occurs only in a few cases, much fewer than in the neutron case, shown in boldface. {This difference can be attributed to the different parameter sets used for protons and for neutrons, as will be discussed in the last paragraph of the present section. } 

b) As a consequence, there are several Nilsson orbitals which have different shell model counterparts for neutrons and for protons. For example, the 1/2[420] Nilsson orbital corresponds to the 1g7/2$_{1/2}$ spherical shell model orbital for neutrons, but to the 2d5/2$_{1/2}$ spherical shell model orbital for protons. More examples of this kind can be found by considering the orbitals appearing in boldface in Table \ref{TA}, but not appearing in boldface in Table \ref{TB}. 

We realize that not only the spin paradox appears in the correspondence between certain asymptotic cylindrical Nilsson orbitals and spherical shell model orbitals, but in addition the correspondence between  asymptotic cylindrical Nilsson orbitals and spherical shell model orbitals is not the same for neutrons and protons. 

It is worth examining the source of these discrepancies by looking at the numerical results of standard Nilsson model calculations, reported in Tables \ref{TD} and \ref{TE}. 

In Table \ref{TD} some examples of proton orbitals (upper part) and neutron orbitals (lower part) are shown. Each Nilsson orbital is expanded in the shell model basis $| N l j \Omega\rangle$. The expansion coefficients are given for three different values of the deformation $\delta$. We remark that in all cases there is a leading contribution by a term having a coefficient very close to unity, shown in boldface. Therefore in these cases there is a clear one-to-one correspondence between the asymptotic cylindrical Nilsson orbitals and the spherical shell model orbitals, as shown in Tables \ref{TA} and \ref{TB}. No spin paradox or mixing of $n_\rho$ or $n_r$ values is observed in these cases. 

Another set of examples of proton orbitals (upper part) and neutron orbitals (lower part) are shown in Table \ref{TE}. In the upper part, we see that there is again a leading term with a large coefficient, shown in boldface, although the dominance of this term gets less absolute for high deformation $\delta$, especially for vectors with low $\Omega$. In other words, high deformation $\delta$ in the case of Nilsson vectors with low $\Omega$ favors non-negligible contributions  for more than one spherical shell model vectors. Anyway, the one-to-one correspondence between cylindrical and sphrical orbitals is preserved in this case, with no spin paradox or mixing of $n_\rho$ or $n_r$ values observed.

A different situation appears in the example shown in the lower part of Table \ref{TE}, which regards neutron orbitals.  In this case it is clear that there is a dominant term, shown in boldface, at low deformation, but the situation is radically different at higher deformations. In most of the cases there is a leading contribution, shown in boldface, but it comes from a spherical vector different from the one appearing at low deformation. 
Non-negligible contributions from more than one spherical vectors become more often in Nilsson vectors with low $\Omega$. In Table \ref{TA} it is clear that the correspondence between Nilsson and spherical orbitals is the one appearing at low deformation, which is different from the one appearing at high deformation. Only for the Nilsson orbital 9/2[514] there is agreement on the leading contribution in low and high deformations. In other words, it is the disagreement between the leading contribution in low and in high deformations in Table \ref{TE} which leads to the spin paradox in Table \ref{TA}, as well as to the mixing of the $n_\rho$ and of the $n_r$ eigenvalues.   

{In relation to the physics origin of the spin paradox, the following comments can be made. Since the expression for the energy eigenvalues of the 3D anisotropic (Nilsson) HO, Eq. (\ref{eigen function}), is the same for both protons and neutrons, the different numerical results are due to the different parameter values used for neutrons and for protons, which can be found in the tables given in \cite{NR,BM,proxy1}. For example, in the neutron shell with $N=82$-126, the parameters used are $\kappa=0.0635$ and $\mu=0.422$, while in the proton shell with $Z=82$-126 the parameters are $\kappa=0.0575$ and $\mu=0.652$ \cite{proxy1}. As a consequence, the coefficient of the spin-orbit term in Eq.  (\ref{Ham}), which is proportional to $\kappa$, differs in the present example between protons and neutrons by about 10\%, but the coefficient of the $\textbf{L}^2$ term, which is proportional to $\kappa\mu$, obtaining the values 0.0268 for neutrons and 0.0375 for protons, differs by 40\%, making the bottom of the proton potential much flatter than the bottom of the neutron potential. As a result, the ``spaghetti'' of Nilsson single particle energy levels obtained for protons and for neutrons is different in each case and the correspondence between single particles states in the spherical case and in the largely deformed (asymptotic) case gets different, when using the sets of quantum numbers described above.} 

In the next section we are going to show how the spin paradox and the disagreement between neutron and proton Nilsson diagrams can be resolved. 

\section{``The rule'' and Nilsson orbitals}

In this section, we discuss how to label  states in a manner different from the traditional approach \cite{NR,FS} presented above, using a {\sl rule}. Following this new labeling, magic numbers and  Nilsson diagrams are reproduced exactly using the asymptotic quantum numbers for all values of the deformation, large and small. 

Let us recall the energy of the 3D isotropic HO in cylindrical and in spherical coordinates. As stated above, it is not important which system of coordinates is employed. Thus, one can expect that the energy of the 3D isotropic HO must be the same both in cylindrical and in spherical coordinates
\begin{align}
E_\text{Cyl.} =&E_{\text{Sph.}
},\nonumber \\
\hbar \omega \left(2 n_\rho + n_z + \Lambda + \frac{3}{2}\right) =& \hbar \omega \left( 2 n_r + L + \frac{3}{2} \right),
\end{align}
which leads to
\begin{align}
2 n_\rho + n_z + \Lambda  =  2 n_r + L.
\end{align}
Since several quantum numbers appear in this equation, this equation does not suffice to provide detailed relations between the cylindrical and spherical quantum numbers.
This gives us the freedom to consider an assumption, namely to equate quantum numbers appearing in this equation with coeffecient equal to one, and separately to equate quantum numbers appearing in this equation with coeffecient equal to two, getting in this way 
\begin{align}
\label{firstrule}
n_r =& n_\rho, \\
\label{secondrule}
L =& n_z + \Lambda. 
\end{align} 
Hereafter we call these arbitrarily assumed equations ``{\sl the rule}''.
These equations show an arbitrary way of mapping the quantum numbers of the spherical coordinates onto the quantum numbers of the  cylindrical coordinates when there is no deformation. Applying Eqs. (\ref{firstrule}) and (\ref{secondrule}) for the correspondence between the cylindrical Nilsson orbitals and the spherical shell model orbitals  we obtain Table \ref{TC}, which is valid for both protons and neutrons and is replacing Table \ref{TA} for neutrons and Table \ref{TB} for protons. On Table \ref{TC} the following remarks can be made. 

a) The spin paradox is resolved. Each bunch of orbitals is characterized by the same spin projection $\Sigma$. For example, the 1g7/2 bunch of orbitals is characterized by spin down, while in Table \ref{TA} it was characterized by a mixture of spin up and spin down orbits.  

b) Each bunch of orbitals is characterized by a common value of $n_\rho$ in cylindrical coordinates and a common value of $n_r$ in spherical coordinates, with $n_\rho=n_r$.  For example, the 1g7/2 bunch of orbitals is characterized by $n_\rho=n_r=0$, while in Table \ref{TA} it corresponds to a mixture of $n_r=0$ and $n_r=1$ orbits.  

c) Table \ref{TC} is valid for both protons and neutrons, while separate tables for neutrons, Table \ref{TA}, and protons, Table \ref{TB}, are obtained within the traditional approach \cite{NR,FS}. 

It should be emphasized that ``the rule'' is an arbitrary assumption, which is not related to any transformation between the spherical and cylindrical bases, as the ones appearing in the literature \cite{Chasman,Davies}.

It is now time to examine how ``the rule''  works within specific numerical examples.


\begin{figure*}[htb]

\includegraphics[width=120mm]{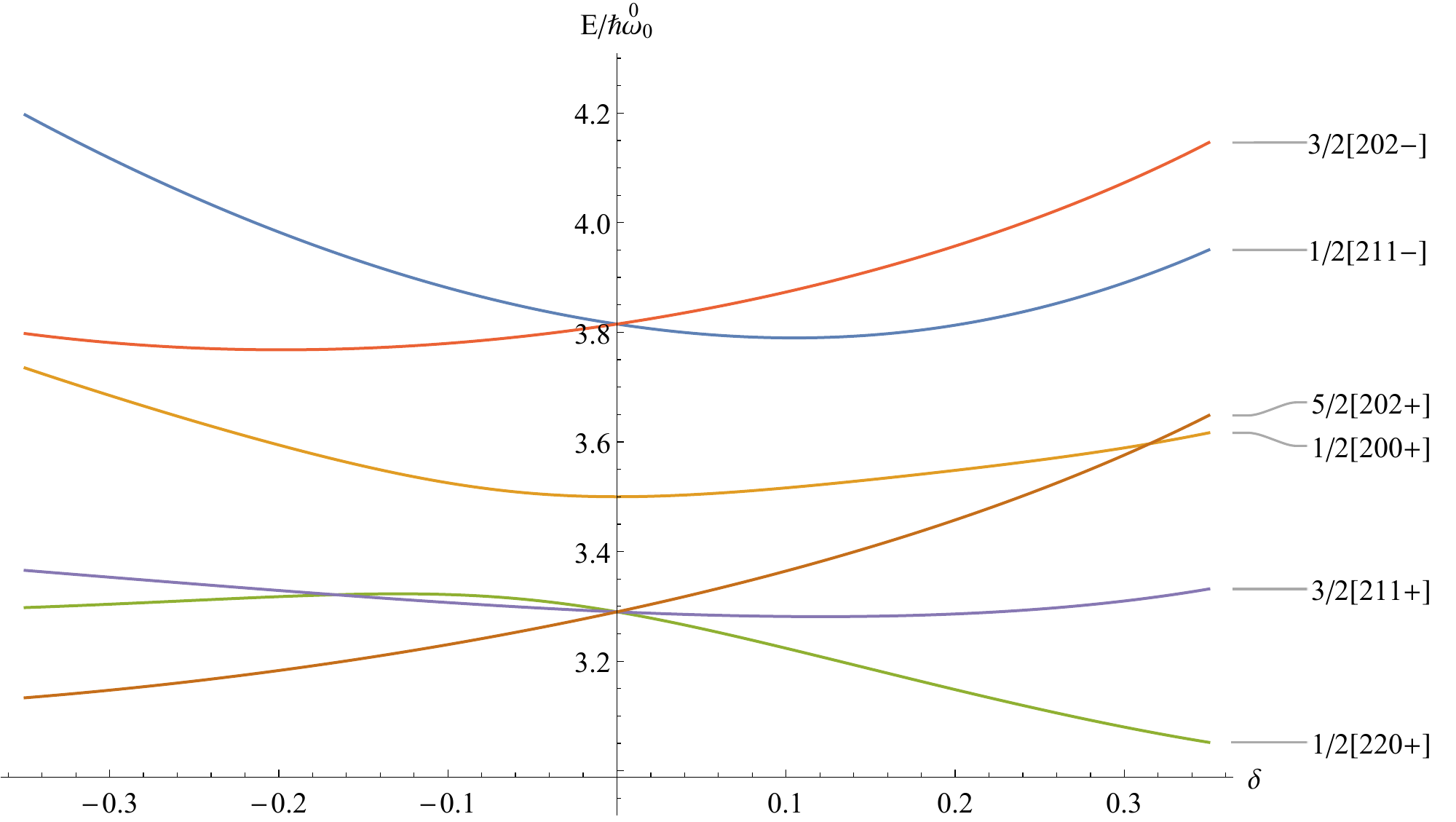}

\caption{Behavior of the levels within the major shell  $ N = 2 $ predicted by ``the rule''. The labels 1/2[200+] and 1/2[211$-$] are interchanged, in comparison to the standard Nilsson diagrams \cite{NR,FS}. 
{The constants are $ \kappa = 0.105 $ and $ \mu = 0 $. This figure holds for both protons and neutrons \cite{NR}.}
} 
\label{FN2}
\end{figure*}


\begin{figure*}[htb]

\includegraphics[width=120mm]{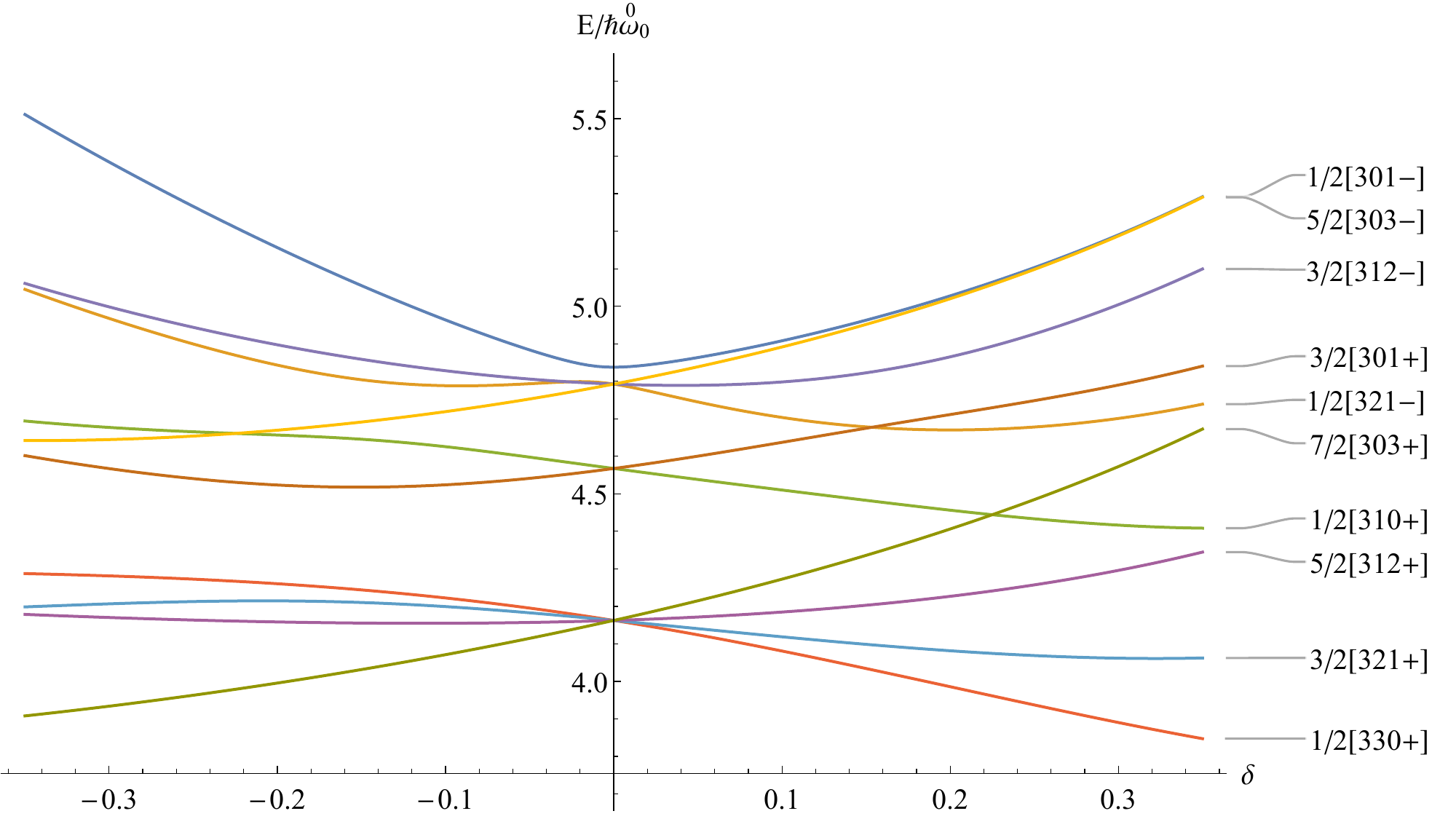}

\caption{The same as Fig. 1, but for $ N = 3 $. The labels 1/2[310+] and 1/2[321$-$] have been interchanged, and in addition the labels 3/2[301+] and 3/2[312$-$] have also been interchanged, in comparison to the standard Nilsson diagrams \cite{NR,FS}. 
{The constants are $ \kappa = 0.090 $ and $ \mu = 0.25 $. This figure holds only for neutrons \cite{NR}.}
} 
\label{FN3}
\end{figure*}

\section{Numerical results}
\label{numerical results}

After determination of the quantum numbers of the various levels according to ``the rule'', as shown in Table \ref{TC}, we now intend to study their behavior as a function of the deformation parameter, $ \delta $, in order  to see if it is possible to use ``the rule'' for plotting Nilsson diagrams. In this direction, we construct explicitly the full Hamiltonian matrix of the modified harmonic oscillator for the shells with $ N = 2$, 3, the results shown in Tables \ref{TN2} and \ref{TN3}. 

In order to obtain the eigenvalues as a function of the deformation parameter $\delta$, these Hamiltonian matrices are diagonalized,  the results of this process plotted in Figs. \ref{FN2} and \ref{FN3}. 

In Fig. \ref{FN2} we see that the lines obtained are identical to these in the standard Nilsson diagrams \cite{NR,FS}, the only difference appearing in the labeling of the states. In particular, the labels 1/2[200+] and 1/2[211$-$] have been interchanged, as expected by comparing Table \ref{TC} to Tables \ref{TA} and \ref{TB}.  

In Fig. \ref{FN3} we see that the lines obtained are identical to these in the standard Nilsson diagrams \cite{NR,FS}, the only difference appearing in the labeling of the states. In particular, the labels 1/2[310+] and 1/2[321$-$] have been interchanged, and in addition the labels 3/2[301+] and 3/2[312$-$] have also been interchanged, as expected by comparing Table \ref{TC} to Tables \ref{TA} and \ref{TB}. 

Let us give an illustrative example of  the above considerations. The spherical shell model wave functions $| N L j \Omega\rangle$ can be expanded in terms of $|N L \Lambda\Sigma\rangle$ wave functions, which explicitly contain $\Sigma$, as follows
\begin{equation}
\label{the expansion}
\left|N, L, j= l \pm \tfrac{1}{2}, \Omega \right>
= \sum _{\Sigma} \left. \left<l \Lambda \tfrac{1}{2} \Sigma \right| j \Omega \right> \left|N L \Lambda \Sigma \right>,
\end{equation}
where the symbol $ \left. \left<l \Lambda \tfrac{1}{2} \Sigma \right| j \Omega \right> $ stands for the appropriate Clebsch-Gordan coefficient, and there is no summation over $\Lambda$ since $\Omega=\Lambda+\Sigma$. Consider, for example,  the state $ | n L j_\Omega\rangle = |1 \text{g} 7/2_{5/2}\rangle $.  Expanding this state according to Eq. \eqref{the expansion} results in 
\begin{align}
\left| 1 \text{g} 7/2 _{5/2} \right> =& 
\left< 4 2 \tfrac{1}{2} \tfrac{1}{2} | \tfrac{7}{2} \tfrac{5}{2} \right> \left| 4 4 2 + \right> + 
 \left< 4 3 \tfrac{1}{2} -\tfrac{1}{2} | \tfrac{7}{2} \tfrac{5}{2} \right>  \left| 4 4 3 - \right>  \nonumber \\
=& -\frac{\sqrt{2}}{3}  \left| 4 4 2 + \right>  + \frac{\sqrt{7}}{3}  \left| 4 4 3 - \right>.
\end{align}
According to the above expression the state contains non-zero
probabilities of both $\Sigma = +1/2$ and $-1/2$ components, with the $\Sigma = -1/2$ component having a larger probability. Since $j = 4 - 1/2$ in the $ 1\text{g}7/2 $ orbital, in a classical picture only the $\Sigma = -1/2$ term might be expected to appear, while according to Table \ref{TA} and Ref. \cite{FS}, for the considered state we have 
\begin{align}
\left| 1 \text{g} 7/2 _{5/2} \right> = 
\begin{cases}
5/2  \left[402+\right] & \text{for neutrons},\\
5/2 \left[413-\right] & \text{for protons}.\\
\end{cases}
\end{align}
In order to find where the paradox occurs, let us calculate the quantum number 
{
$ n_x + n_y \equiv n_\perp  = N - n_z = 2 n_\rho + |\Lambda| $, 
where the last equation is obtained by equating Eqs. \eqref{E-cart} and \eqref{E-cyl}}
needed in the Nilsson Hamiltonian using the asymptotic quantum numbers. 
{ One may check at this point that the same constraint for $ n_\perp $ has been obtained using a different method in Ref. \cite{NR} page 115. }
In the neutron case we have
\begin{align}
5/2 \left[402+\right] \rightarrow 
n_{\perp} = 
\begin{cases}
N - n_z = 4 - 0 = 4,\\
2n_\rho + \Lambda = 0 + 2 = 2,
\end{cases}
\end{align}
while for the proton case we obtain
\begin{align}
5/2 \left[413-\right] \rightarrow 
n_{\perp} = 
\begin{cases}
N - n_z = 4 - 1 = 3,\\
2n_\rho + \Lambda = 0 + 3 = 3.
\end{cases}
\end{align}
It is seen that in the neutron case we cannot reach the same value of $ n_\perp $, while we do  obtain  the same value of $ n_\perp $ in the proton case, which follows ``the rule''.  

In other words, since the Clebsch-Gordan coefficients are the same for both protons and neutrons, the same results would have been expected in both cases. Although in the classical picture in $ 1\text{g}7/2 $ one expects $ \Sigma=-1/2 $, this is indeed the case for protons, shown in Table \ref{TB}, but not for neutrons, shown in Table \ref{TA}. This is what we call a paradox. ``The rule'' gives the right answer. The correspondence under discussion is valid only for protons, i.e., in the case in which ``the rule'' is satisfied.

We conclude that Nilsson diagrams can be exactly reproduced following ``the rule''. The single-particle energy levels as a function of the quadrupole deformation $\delta$ look exactly the same as in the traditional approach \cite{NR,FS}, the only difference being that the Nilsson labels for certain pairs of energy levels are interchanged. These interchanges resolve the spin paradox seen in the traditional Nilsson diagrams, and in addition they make the Nilsson diagrams for protons and for neutrons identical to each other, while the traditional Nilsson diagrams for protons and for neutrons are different from each other \cite{NR,FS}.  

\section{Conclusion}

In this paper a new way of establishing the correspondence between the quantum numbers characterizing single-particle nuclear levels in the spherical coordinates used in the shell model and in the cylindrical coordinates used in the Nilsson model is introduced. The new correspondence resolves the spin paradox seen in the traditional approach, since in the new description each bunch of levels characterized by the same orbital angular momentum $l$ and the same total angular momentum $j$ in the spherical coordinates corresponds to a single value of the spin projection in the cylindrical coordinates, and not to a mixture of spin projections, like in the traditional description. Furthermore, the correspondence between spherical shell model orbitals and asymptotic cylindrical Nilsson orbitals becomes the same for both protons and neutrons, while different sets of correspondence are obtained for protons and for neutrons within the traditional approach. Numerical results prove that the dependence of the single-particle energy levels on the quadrupole deformation, shown in the  Nilsson diagrams, remains exactly the same as in the traditional approach, the only difference being that for certain pairs of single particle energy levels the Nilsson labels corresponding to two spherical shell model orbitals get interchanged.

{
\textbf{Acknowledgement}:
The authors are grateful to the referees for helpful comments and suggestions
on this work.
}

\begin{table*}[htb]

\caption{Correspondence between Nilsson orbitals $\Omega[N n_z \Lambda]$ and shell model orbitals $n l j_\Omega$, as given in the standard Nilsson diagrams for neutrons \cite{FS}. The quantum numbers $n_\rho$ and $\Sigma=\Omega-\Lambda$ in the cylindrical coordinates used by the Nilsson model and $n_r=n-1$ in the spherical coordinates used by the shell model are also shown. See Section 4 for further discussion. Spherical orbitals involved in the discussion are shown in boldface. Below the name of each shell the orbitals belonging to it appear. }
\label{TA}       
\vskip 0.5cm 
\begin{tabular}{ r r r r r| r r r r r| r r r r r   }

$\Omega[N n_z \Lambda]$ & $n_\rho$ & $\Sigma$ & $nlj_\Omega$ & $n_r$ & 
$\Omega[N n_z \Lambda]$ & $n_\rho$ & $\Sigma$ & $nlj_\Omega$ & $n_r$ & 
$\Omega[N n_z \Lambda]$ & $n_\rho$ & $\Sigma$ & $nlj_\Omega$ & $n_r$ \\

\hline\noalign{\smallskip}
 sdg  & & & &  & pfh  & & & &  & sdgi & & & & \\ 
1/2[440] &0&+& 1g9/2$_{1/2}$ &0& 1/2[550] &0&+&1h11/2$_{1/2}$ &0& 1/2[660] &0&+&1i13/2$_{1/2}$ & 0\\
3/2[431] &0&+& 1g9/2$_{3/2}$ &0& 3/2[541] &0&+&1h11/2$_{3/2}$ &0& 3/2[651] &0&+&1i13/2$_{3/2}$ & 0\\
5/2[422] &0&+& 1g9/2$_{5/2}$ &0& 5/2[532] &0&+&1h11/2$_{5/2}$ &0& 5/2[642] &0&+&1i13/2$_{5/2}$ & 0\\
7/2[413] &0&+& 1g9/2$_{7/2}$ &0& 7/2[523] &0&+&1h11/2$_{7/2}$ &0& 7/2[633] &0&+&1i13/2$_{7/2}$ & 0\\
9/2[404] &0&+& 1g9/2$_{9/2}$ &0& 9/2[514] &0&+&1h11/2$_{9/2}$ &0& 9/2[624] &0&+&1i13/2$_{9/2}$ & 0\\
         & & &            & &11/2[505] &0&+&1h11/2$_{11/2}$ &0&11/2[615] &0&+&1i13/2$_{11/2}$ & 0\\
1/2[431] &0&$-$&{\bf 2d5/2}$_{1/2}$ &1&          & & &          & &13/2[606] &0&+&1i13/2$_{13/2}$ & 0\\
3/2[422] &0&$-$&{\bf 2d5/2}$_{3/2}$ &1& 1/2[541] &0&$-$&{\bf 2f7/2}$_{1/2}$ &1&          & & &       &  \\
5/2[413] &0&$-$&{\bf 2d5/2}$_{5/2}$ &1& 3/2[532] &0&$-$&{\bf 2f7/2}$_{3/2}$ &1& 1/2[651] &0&$-$&{\bf 2g9/2}$_{1/2}$ &1 \\
7/2[404] &0&$-$& 1g7/2$_{7/2}$ &0& 5/2[523] &0&$-$&{\bf 2f7/2}$_{5/2}$ &1& 3/2[642] &0&$-$&{\bf 2g9/2}$_{3/2}$ &1 \\
         & &   &            & & 7/2[514] &0&$-$&{\bf 2f7/2}$_{7/2}$ &1& 5/2[633] &0&$-$&{\bf 2g9/2}$_{5/2}$ &1 \\
1/2[420] &1&+&{\bf 1g7/2}$_{1/2}$ &0& 9/2[505] &0&$-$& 1h9/2$_{9/2}$ &0& 7/2[624] &0&$-$&{\bf 2g9/2}$_{7/2}$ &1 \\
3/2[411] &1&+&{\bf 1g7/2}$_{3/2}$ &0&          & &   &            & & 9/2[615] &0&$-$&{\bf 2g9/2}$_{9/2}$ &1 \\
5/2[402] &1&+&{\bf 1g7/2}$_{5/2}$ &0& 1/2[530] &1&+&{\bf 1h9/2}$_{1/2}$ &0&11/2[606] &0&$-$&1i11/2$_{11/2}$ &0 \\
         & & &            & & 3/2[521] &1&+&{\bf 1h9/2}$_{3/2}$ &0&          & &   &            & \\
1/2[411] &1&$-$&{\bf 3s1/2}$_{1/2}$ &2& 5/2[512] &1&+&{\bf 1h9/2}$_{5/2}$ &0& 1/2[640] &1&+&{\bf 1i11/2}$_{1/2}$ &0 \\
3/2[402] &1&$-$& 2d3/2$_{3/2}$ &1& 7/2[503] &1&+&{\bf 1h9/2}$_{7/2}$ &0& 3/2[631] &1&+&{\bf 1i11/2}$_{3/2}$ &0 \\
          & & &       & &          & & &       &                          & 5/2[622] & 1&+&{\bf 1i11/2}$_{5/2}$ &0 \\ 
1/2[400] &2&+&{\bf 2d3/2}$_{1/2}$ &1& 1/2[521] &1&$-$&{\bf 3p3/2}$_{1/2}$ &2& 7/2[613] &1&+&{\bf 1i11/2}$_{7/2}$ &0 \\
          & & &       & & 3/2[512] &1&$-$&{\bf 3p3/2}$_{3/2}$ &2& 9/2[604] &1&+&{\bf 1i11/2}$_{9/2}$ &0 \\
   s    & & &       & & 5/2[503] &1&$-$& 2f5/2$_{5/2}$ &1&          & & &       & \\
1/2[000] &0&+& 1s1/2$_{1/2}$ &0&           & & &       & & 1/2[631] &1&$-$&{\bf 3d5/2}$_{1/2}$ &2 \\
          & & &       & & 1/2[510] &2&+&{\bf 2f5/2}$_{1/2}$ &1& 3/2[622] &1&$-$&{\bf 3d5/2}$_{3/2}$ &2 \\
 p         & & &       & & 3/2[501] &2&+&{\bf 2f5/2}$_{3/2}$ &1& 5/2[613] &1&$-$&{\bf 3d5/2}$_{5/2}$ &2 \\
 1/2[110] &0&+& 1p3/2$_{1/2}$ &0&               & & &       & & 7/2[604] &1&$-$& 2g7/2$_{7/2}$ &1 \\
 3/2[101] &0&+& 1p3/2$_{3/2}$ &0& 1/2[501] &2&$-$& 3p1/2$_{1/2}$ &2&          & & &            & \\
          & & &            & &          & &   &            & & 1/2[620] &2&+&{\bf 2g7/2}$_{1/2}$ &1 \\
1/2[101] &0&$-$& 1p1/2$_{1/2}$ &0&  pf           & & &       & & 3/2[611] &2&+&{\bf 2g7/2}$_{3/2}$ &1 \\
         & &  &             & &1/2[330] &0&+& 1f7/2$_{1/2}$ &0               & 5/2[602] &2&+&{\bf  2g7/2}$_{5/2}$ &1 \\
 sd      & &  &             & &3/2[321] &0&+& 1f7/2$_{3/2}$ &0                &          & & &       & \\
1/2[220] &0& +& 1d5/2$_{1/2}$ &0&5/2[312] &0&+& 1f7/2$_{5/2}$ &0               & 1/2[611] &2&$-$&{\bf 4s1/2}$_{1/2}$ &3 \\      
3/2[211] &0& +& 1d5/2$_{3/2}$ &0&7/2[303] &0&+& 1f7/2$_{7/2}$ &0  &     3/2[602] &2&$-$& 3d3/2$_{3/2}$ &2 \\
5/2[202] &0& +& 1d5/2$_{5/2}$ &0&                & & &       & &              & &   &            & \\
         & &  &            & &1/2[321] &0&$-$&{\bf 2p3/2}$_{1/2}$ &1          & 1/2[600] &3&+&{\bf 3d3/2}$_{1/2}$ &2 \\
1/2[211]&0&$-$&{\bf 2s1/2}$_{1/2}$ &1&3/2[312] &0&$-$&{\bf 2p3/2}$_{3/2}$ &1&  & & & & \\
3/2[202]&0&$-$& 1d3/2$_{3/2}$ &0&5/2[303] &0&$-$& 1f5/2$_{5/2}$ &0&   & & & & \\
         & & &             & &         & &   &            & &    & & & & \\ 
1/2[200] &1&+&{\bf 1d3/2}$_{1/2}$  &0&1/2[310] &1&+&{\bf 1f5/2}$_{1/2}$ &0& & & & & \\
         & & &             & &3/2[301] &1&+&{\bf 1f5/2}$_{3/2}$ &0&    & & & & \\
         & & &             & &          & & &       & &         & & & & \\
         & & &             & &1/2[301] &1&$-$& 2p1/2$_{1/2}$ &1&  & & & & \\

\end{tabular}
\end{table*}

\begin{table*}[htb]

\caption{Correspondence between Nilsson orbitals $\Omega[N n_z \Lambda]$  and shell model orbitals $n l^j_\Omega$, as given in the standard Nilsson diagrams for protons\cite{FS}. The quantum numbers $n_\rho$ and $\Sigma=\Omega-\Lambda$ in the cylindrical coordinates used by the Nilsson model and $n_r$ in the spherical coordinates used by the shell model are also shown. See Section 4 for further discussion. Spherical orbitals involved in the discussion are shown in boldface. Below the name of each shell the orbitals belonging to it appear. }
\label{TB}       
\vskip 0.5cm 
\begin{tabular}{ r r r r r| r r r r r    }

$\Omega[N n_z \Lambda]$ & $n_\rho$ & $\Sigma$ & $nlj_\Omega$ & $n_r$ & 
$\Omega[N n_z \Lambda]$ & $n_\rho$ & $\Sigma$ & $nlj_\Omega$ & $n_r$ \\

\hline\noalign{\smallskip} 
 sdg  & & & &  & pfh  & & & &    \\ 
1/2[440] &0&+& 1g9/2$_{1/2}$ &0& 1/2[550] &0&+&1h11/2$_{1/2}$ &0\\
3/2[431] &0&+& 1g9/2$_{3/2}$ &0& 3/2[541] &0&+&1h11/2$_{3/2}$ &0\\
5/2[422] &0&+& 1g9/2$_{5/2}$ &0& 5/2[532] &0&+&1h11/2$_{5/2}$ &0\\
7/2[413] &0&+& 1g9/2$_{7/2}$ &0& 7/2[523] &0&+&1h11/2$_{7/2}$ &0\\
9/2[404] &0&+& 1g9/2$_{9/2}$ &0& 9/2[514] &0&+&1h11/2$_{9/2}$ &0\\
         & & &            & &11/2[505] &0&+&1h11/2$_{11/2}$ &0\\
1/2[431] &0&$-$& 1g7/2$_{1/2}$ &0&          & & &          & \\
3/2[422] &0&$-$& 1g7/2$_{3/2}$ &0& 1/2[541] &0&$-$& 1h9/2$_{1/2}$ &0\\   
5/2[413] &0&$-$& 1g7/2$_{5/2}$ &0& 3/2[532] &0&$-$& 1h9/2$_{3/2}$ &0\\
7/2[404] &0&$-$& 1g7/2$_{7/2}$ &0& 5/2[523] &0&$-$& 1h9/2$_{5/2}$ &0\\
         & &   &            & & 7/2[514] &0&$-$& 1h9/2$_{7/2}$ &0\\
1/2[420] &1&+& 2d5/2$_{1/2}$ &1& 9/2[505] &0&$-$& 1h9/2$_{9/2}$ &0\\
3/2[411] &1&+& 2d5/2$_{3/2}$ &1&          & &   &            & \\
5/2[402] &1&+& 2d5/2$_{5/2}$ &1& 1/2[530] &1&+& 2f7/2$_{1/2}$ &1\\
         & & &            & & 3/2[521] &1&+& 2f7/2$_{3/2}$ &1\\
1/2[411] &1&$-$& 2d3/2$_{1/2}$ &1& 5/2[512] &1&+& 2f7/2$_{5/2}$ &1\\
3/2[402] &1&$-$& 2d3/2$_{3/2}$ &1& 7/2[503] &1&+& 2f7/2$_{7/2}$ &1\\
          & & &       & &          & & &       & \\
1/2[400] &2&+& 3s1/2$_{1/2}$ &2& 1/2[521] &1&$-$& 2f5/2$_{1/2}$ &1\\ 
          & & &       & & 3/2[512] &1&$-$& 2f5/2$_{3/2}$ &1\\
   s    & & &       & & 5/2[503] &1&$-$& 2f5/2$_{5/2}$ &1\\
1/2[000] &0&+& 1s1/2$_{1/2}$ &0&           & & &       & \\
          & & &       & & 1/2[510] &2&+& 3p3/2$_{1/2}$ &2\\
 p         & & &       & & 3/2[501] &2&+& 3p3/2$_{3/2}$ &2\\
 1/2[110] &0&+& 1p3/2$_{1/2}$ &0&               & & &       & \\ 
 3/2[101] &0&+& 1p3/2$_{3/2}$ &0& 1/2[501] &2&$-$& 3p1/2$_{1/2}$ &2\\  
          & & &            & &          & &   &            & \\ 
1/2[101] &0&$-$& 1p1/2$_{1/2}$ &0&  pf           & & &       & \\
         & &  &             & &1/2[330] &0&+& 1f7/2$_{1/2}$ &0\\ 
 sd      & &  &             & &3/2[321] &0&+& 1f7/2$_{3/2}$ &0 \\      
1/2[220] &0& +& 1d5/2$_{1/2}$ &0&5/2[312] &0&+& 1f7/2$_{5/2}$ &0 \\    
3/2[211] &0& +& 1d5/2$_{3/2}$ &0&7/2[303] &0&+& 1f7/2$_{7/2}$ &0  \\     
5/2[202] &0& +& 1d5/2$_{5/2}$ &0&                & & &       & \\           
         & &  &            & &1/2[321] &0&$-$&{\bf 2p3/2}$_{1/2}$ &1\\
1/2[211]&0&$-$&{\bf 2s1/2}$_{1/2}$ &1&3/2[312]   &0&$-$&{\bf 2p3/2}$_{3/2}$ &1\\
3/2[202]&0&$-$& 1d3/2$_{3/2}$      &0&5/2[303] &0&$-$& 1f5/2$_{5/2}$ &0\\ 
         & & &             & &         & &   &            & \\  
1/2[200] &1&+&{\bf 1d3/2}$_{1/2}$  &0&1/2[310] &1&+&{\bf 1f5/2}$_{1/2}$ &0\\
         & & &             & &3/2[301] &1&+&{\bf 1f5/2}$_{3/2}$ &0\\  
         & & &             & &          & & &       & \\      
         & & &             & &1/2[301] &1&$-$& 2p1/2$_{1/2}$ &1\\

\end{tabular}
\end{table*}

\begin{table*}[htb]

\caption{Expansions of Nilsson orbitals $\Omega[N n_z \Lambda]$ in the shell model basis $|N l j \Omega \rangle$ for three different values of the deformation $\delta$. In the upper part, results in the 50-82 proton shell are shown, obtained with the parameter values $\kappa= 0.0637$ and $\mu=0.60$,  while in the lower part, results in the 82-126 neutron shell are shown, obtained with the parameter values $\kappa= 0.0637$ and $\mu=0.42$.
The existence of a leading shell model eigenvector is evident at all deformations. See Section 4 for further discussion. 
}
\label{TD}       
\vskip 0.5cm 
\begin{tabular}{ r r r r r r r r }

$\Omega[N n_z \Lambda]$ & $\delta$ & & & & & & \\

\hline\noalign{\smallskip}
${1\over 2}[550]$ &  &
$\left| 5 1 {1\over 2} {1\over 2} \right\rangle$ & 
$\left| 5 1 {3\over 2} {1\over 2} \right\rangle$ & 
$\left| 5 3 {5\over 2} {1\over 2} \right\rangle$ & 
$\left| 5 3 {7\over 2} {1\over 2} \right\rangle$ & 
$\left| 5 5 {9\over 2} {1\over 2} \right\rangle$ &
$\left| 5 5 {11\over 2} {1\over 2} \right\rangle$ \\
&0.05 & $-0.0002$ & 0.0039 & $-0.0006$ & 0.0709 & $-0.0043$ & {\bf 0.9975} \\
&0.22 & $-0.0132$ & 0.0692 & $-0.0179$ & 0.2966 & $-0.0258$ & {\bf 0.9519} \\
&0.30 & $-0.0306$ & 0.1207 & $-0.0381$ & 0.3867 & $-0.0418$ & {\bf 0.9120} \\

\noalign{\smallskip}\hline\noalign{\smallskip}
${3\over 2}[541]$ &  & &
$\left| 5 1 {3\over 2} {3\over 2} \right\rangle$ & 
$\left| 5 3 {5\over 2} {3\over 2} \right\rangle$ & 
$\left| 5 3 {7\over 2} {3\over 2} \right\rangle$ & 
$\left| 5 5 {9\over 2} {3\over 2} \right\rangle$ & 
$\left| 5 5 {11\over 2} {3\over 2} \right\rangle$ \\
&0.05 & & 0.0025 & $-0.0015$ & 0.0641 & $-0.0122$ & {\bf 0.9979} \\
&0.22 & & 0.0371 & $-0.0286$ & 0.2565 & $-0.0640$ & {\bf 0.9633} \\
&0.30 & & 0.0601 & $-0.0506$ & 0.3287 & $-0.0922$ & {\bf 0.9366} \\

\noalign{\smallskip}\hline\noalign{\smallskip}
${5\over 2}[532]$ &  & & &
$\left| 5 3 {5\over 2} {5\over 2} \right\rangle$ & 
$\left| 5 3 {7\over 2} {5\over 2} \right\rangle$ & 
$\left| 5 5 {9\over 2} {5\over 2} \right\rangle$ & 
$\left| 5 5 {11\over 2} {5\over 2} \right\rangle$ \\
&0.05 & & & $-0.0014$ & 0.0511 & $-0.0182$ & {\bf 0.9985} \\
&0.22 & & & $-0.0202$ & 0.1915 & $-0.0842$ & {\bf 0.9777} \\
&0.30 & & & $-0.0325$ & 0.2411 & $-0.1144$ & {\bf 0.9632} \\

\noalign{\smallskip}\hline\noalign{\smallskip}
${7\over 2}[523]$ &  & & & &
$\left| 5 3 {7\over 2} {7\over 2} \right\rangle$ & 
$\left| 5 5 {9\over 2} {7\over 2} \right\rangle$ & 
$\left| 5 5 {11\over 2} {7\over 2} \right\rangle$ \\
&0.05 & & & & 0.0323 & $-0.0212$ & {\bf 0.9993} \\ 
&0.22 & & & & 0.1129 & $-0.0872$ & {\bf 0.9898} \\
&0.30 & & & & 0.1398 & $-0.1138$ & {\bf 0.9836} \\

\noalign{\smallskip}\hline\noalign{\smallskip}
${9\over 2}[514]$ &  & & & & &
$\left| 5 5 {9\over 2} {9\over 2} \right\rangle$ & 
$\left| 5 5 {11\over 2} {9\over 2} \right\rangle$ \\
&0.05 & & & & & $-0.0194$ & {\bf 0.9998} \\  
&0.22 & & & & & $-0.0716$ & {\bf 0.9974} \\
&0.30 & & & & & $-0.0907$ & {\bf 0.9959} \\

\noalign{\smallskip}\hline\noalign{\smallskip}
${11\over 2}[505]$ &  & & & & & & 
$\left| 5 5 {11\over 2} {11\over 2} \right\rangle$ \\
&0.05 & & & & & & {\bf 1.0000} \\  
&0.22 & & & & & & {\bf 1.0000} \\
&0.30 & & & & & & {\bf 1.0000} \\

\end{tabular}

\vskip 0.5cm 

\begin{tabular}{ r r r r r r r r r }

$\Omega[N n_z \Lambda]$ & $\delta$ & & & & & & & \\

\hline\noalign{\smallskip}
${1\over 2}[660]$ & &
$\left| 6 0 {1\over 2} {1\over 2} \right\rangle$ & 
$\left| 6 2 {3\over 2} {1\over 2} \right\rangle$ & 
$\left| 6 2 {5\over 2} {1\over 2} \right\rangle$ & 
$\left| 6 4 {7\over 2} {1\over 2} \right\rangle$ & 
$\left| 6 4 {9\over 2} {1\over 2} \right\rangle$ & 
$\left| 6 6 {11\over 2} {1\over 2} \right\rangle$ &
$\left| 6 6 {13\over 2} {1\over 2} \right\rangle$ \\
&0.05 & 0.0003 & $-0.0001$ & 0.0059 & $-0.0005$ & 0.0879 & $-0.0030$ & {\bf 0.9961} \\ 
&0.22 & 0.0225 & $-0.0118$ & 0.1035 & $-0.0144$ & 0.3589 & $-0.0179$ & {\bf 0.9270} \\
&0.30 & 0.0500 & $-0.0305$ & 0.1772 & $-0.0314$ & 0.4571 & $-0.0290$ & {\bf 0.8686} \\ 

\noalign{\smallskip}\hline\noalign{\smallskip}
${3\over 2}[651]$ & & &
$\left| 6 2 {3\over 2} {3\over 2} \right\rangle$ & 
$\left| 6 2 {5\over 2} {3\over 2} \right\rangle$ & 
$\left| 6 4 {7\over 2} {3\over 2} \right\rangle$ & 
$\left| 6 4 {9\over 2} {3\over 2} \right\rangle$ & 
$\left| 6 6 {11\over 2} {3\over 2} \right\rangle$ &
$\left| 6 6 {13\over 2} {3\over 2} \right\rangle$ \\
&0.05 & & $-0.0002$ & 0.0046 & $-0.0013$ & 0.0821 & $-0.0086$ & {\bf 0.9966}  \\
&0.22 & & $-0.0100$ & 0.0711 & $-0.0278$ & 0.3240 & $-0.0469$ & {\bf 0.9418}  \\
&0.30 & & $-0.0207$ & 0.1149 & $-0.0509$ & 0.4091 & $-0.0687$ & {\bf 0.9010}  \\

\noalign{\smallskip}\hline\noalign{\smallskip}
${5\over 2}[642]$ & & & & 
$\left| 6 2 {5\over 2} {5\over 2} \right\rangle$ & 
$\left| 6 4 {7\over 2} {5\over 2} \right\rangle$ & 
$\left| 6 4 {9\over 2} {5\over 2} \right\rangle$ & 
$\left| 6 6 {11\over 2} {5\over 2} \right\rangle$ &
$\left| 6 6 {13\over 2} {5\over 2} \right\rangle$ \\
&0.05 & & & 0.0026 & $-0.0016$ & 0.0707 & $-0.0134$ & {\bf 0.9974} \\
&0.22 & & & 0.0337 & $-0.0262$ & 0.2654 & $-0.0658$ & {\bf 0.9610} \\
&0.30 & & & 0.0519 & $-0.0438$ & 0.3315 & $-0.0917$ & {\bf 0.9365} \\

\noalign{\smallskip}\hline\noalign{\smallskip}
${7\over 2}[633]$ & & & & &
$\left| 6 4 {7\over 2} {7\over 2} \right\rangle$ & 
$\left| 6 4 {9\over 2} {7\over 2} \right\rangle$ & 
$\left| 6 6 {11\over 2} {7\over 2} \right\rangle$ &
$\left| 6 6 {13\over 2} {7\over 2} \right\rangle$ \\
&0.05 & & & & $-0.0012$ & 0.0544 & $-0.0166$ & {\bf 0.9984} \\
&0.22 & & & & $-0.0161$ & 0.1927 & $-0.0744$ & {\bf 0.9783} \\
&0.30 & & & & $-0.0254$ & 0.2380 & $-0.1000$ & {\bf 0.9658} \\

\noalign{\smallskip}\hline\noalign{\smallskip}
${9\over 2}[624]$ & & & & & & 
$\left| 6 4 {9\over 2} {9\over 2} \right\rangle$ & 
$\left| 6 6 {11\over 2} {9\over 2} \right\rangle$ &
$\left| 6 6 {13\over 2} {9\over 2} \right\rangle$ \\
&0.05 & & & & & 0.0336 & $-0.0176$ & {\bf 0.9993} \\
&0.22 & & & & & 0.1120 & $-0.0724$ & {\bf 0.9911} \\
&0.30 & & & & & 0.1366 & $-0.0947$ & {\bf 0.9861} \\

\noalign{\smallskip}\hline\noalign{\smallskip}
${11\over 2}[615]$ & & & & & & &
$\left| 6 6 {11\over 2} {11\over 2} \right\rangle$ &
$\left| 6 6 {13\over 2} {11\over 2} \right\rangle$ \\
&0.05 & & & & & & $-0.0153$ & {\bf 0.9999} \\
&0.22 & & & & & & $-0.0578$ & {\bf 0.9983} \\
&0.30 & & & & & & $-0.0739$ & {\bf 0.9973} \\

\noalign{\smallskip}\hline\noalign{\smallskip}
${13\over 2}[606]$ & & & & & & & &
$\left| 6 6 {13\over 2} {13\over 2} \right\rangle$ \\
&0.05 & & & & & & & {\bf 1.000} \\
&0.22 & & & & & & & {\bf 1.000} \\
&0.30 & & & & & & & {\bf 1.000} \\

\end{tabular}

\end{table*}

\begin{table*}[htb]

\caption{Expansions of Nilsson orbitals $\Omega[N n_z \Lambda]$ in the shell model basis $|N l j \Omega \rangle$ for three different values of the deformation $\delta$. In the upper part, results in the 50-82 proton shell are shown, obtained with the parameter values $\kappa= 0.0637$ and $\mu=0.60$,  while in the lower part, results in the 82-126 neutron shell are shown, obtained with the parameter values $\kappa= 0.0637$ and $\mu=0.42$. The existence of a leading shell model eigenvector is evident at small deformation, but this is not the case any more at higher deformations, at which several shell model eigenvectors make considerable contributions. See Section 4 for further discussion.   
}
\label{TE}       

\vskip 0.5cm 
\begin{tabular}{ r r r r r r r }

$\Omega[N n_z \Lambda]$&  $\delta$ & & & & & \\

\hline\noalign{\smallskip} 

${1\over 2}[431]$ & & $\left| 4 0 {1\over 2} {1\over 2} \right\rangle$ & $\left| 4 2 {3\over 2} {1\over 2} \right\rangle$ & 
$\left| 4 2 {5\over 2} {1\over 2} \right\rangle$ & $\left| 4 4 {7\over 2} {1\over 2} \right\rangle$ & $\left| 4 4 {9\over 2} {1\over 2} \right\rangle$ \\

&0.05 & $-0.0213$ & 0.1254 & $-0.0702$ & {\bf 0.9893} & 0.0127 \\
&0.22 & $-0.2248$ & 0.4393 & $-0.2791$ & {\bf 0.8057} & 0.1717 \\
&0.30 & $-0.2630$ & 0.5003 & $-0.2458$ & {\bf 0.7447} & 0.2559 \\

\noalign{\smallskip}\hline\noalign{\smallskip}

                               ${3\over 2}[422]$ & &  & $\left| 4 2 {3\over 2} {3\over 2} \right\rangle$ & 
$\left| 4 2 {5\over 2} {3\over 2} \right\rangle$ & $\left| 4 4 {7\over 2} {3\over 2} \right\rangle$ & $\left| 4 4 {9\over 2} {3\over 2} \right\rangle$ \\

&0.05 & & 0.0882 & $-0.1290$ & {\bf 0.9873} & 0.0275 \\
&0.22 & & 0.2576 & $-0.2540$ & {\bf 0.9172} & 0.1669 \\
&0.30 & & 0.2925 & $-0.2396$ & {\bf 0.8979} & 0.2253 \\

\noalign{\smallskip}\hline\noalign{\smallskip}

                               ${5\over 2}[413]$ & &  & & 
$\left| 4 2 {5\over 2} {5\over 2} \right\rangle$ & $\left| 4 4 {7\over 2} {5\over 2} \right\rangle$ & $\left| 4 4 {9\over 2} {5\over 2} \right\rangle$ \\

&0.05 & &  & $-0.1149$ & {\bf 0.9929} & 0.0309 \\
&0.22 & &  & $-0.1832$ & {\bf 0.9737} & 0.1356 \\
&0.30 & &  & $-0.1800$ & {\bf 0.9680} & 0.1748 \\

\noalign{\smallskip}\hline\noalign{\smallskip}

                               ${7\over 2}[404]$ & &  & & 
 & $\left| 4 4 {7\over 2} {7\over 2} \right\rangle$ & $\left| 4 4 {9\over 2} {7\over 2} \right\rangle$ \\

&0.05 & &  &  & {\bf 0.9997} & 0.0256 \\
&0.22 & &  &  & {\bf 0.9958} & 0.0921 \\
&0.30 & &  &  & {\bf 0.9933} & 0.1154 \\

\end{tabular}
\vskip 0.5cm 

\begin{tabular}{ r r r r r r r r }

$\Omega[N n_z \Lambda]$&  $\delta$ & & & & & & \\
\hline\noalign{\smallskip}

${1\over 2}[541]$ & & 
$\left| 5 1 {1\over 2} {1\over 2} \right\rangle$ & $\left| 5 1 {3\over 2} {1\over 2} \right\rangle$ & $\left| 5 3 {5\over 2} {1\over 2} \right\rangle$ & 
$\left| 5 3 {7\over 2} {1\over 2} \right\rangle$ & $\left| 5 5 {9\over 2} {1\over 2} \right\rangle$ & $\left| 5 5 {11\over 2} {1\over 2} \right\rangle$ \\

&0.05 & $-0.0200$ & 0.1770 & $-0.0295$ & {\bf 0.9780} & $-0.0446$ & $-0.0944$  \\
&0.22 & $-0.2492$ & 0.4619 & $-0.3768$ & 0.5550 & $-0.4161$ & $-0.3185$  \\
&0.30 & $-0.3121$ & 0.4331 & $-0.4829$ & 0.3430 & $-0.4789$ & $-0.3671$  \\

\noalign{\smallskip}\hline\noalign{\smallskip}

${3\over 2}[532]$ & & 
 & $\left| 5 1 {3\over 2} {3\over 2} \right\rangle$ & $\left| 5 3 {5\over 2} {3\over 2} \right\rangle$ & 
$\left| 5 3 {7\over 2} {3\over 2} \right\rangle$ & $\left| 5 5 {9\over 2} {3\over 2} \right\rangle$ & $\left| 5 5 {11\over 2} {3\over 2} \right\rangle$ \\

&0.05 &  & 0.1168    & $-0.0628$ & {\bf 0.9796} & $-0.1245$ & $-0.0860$  \\
&0.22 &  & $-0.2341$ & 0.3692 & $-0.5596$ & {\bf 0.6511} & 0.2681 \\
&0.30 &  & $-0.2237$ & 0.4341  & $-0.4053$ & {\bf 0.7083} & 0.3093  \\

\noalign{\smallskip}\hline\noalign{\smallskip}

${5\over 2}[523]$ & & 
 &  & $\left| 5 3 {5\over 2} {5\over 2} \right\rangle$ & 
$\left| 5 3 {7\over 2} {5\over 2} \right\rangle$ & $\left| 5 5 {9\over 2} {5\over 2} \right\rangle$ & $\left| 5 5 {11\over 2} {5\over 2} \right\rangle$ \\

&0.05 &  &  & $-0.0630$ & {\bf 0.9759} & $-0.1968$ & $-0.0696$  \\
&0.22 &  &  & 0.2371 & $-0.4724$ & {\bf 0.8260} & 0.1959 \\
&0.30 &  &  & 0.2648  & $-0.3657$ & {\bf 0.8622} & 0.2297  \\

\noalign{\smallskip}\hline\noalign{\smallskip}

${7\over 2}[514]$ & & 
 &  &  & 
$\left| 5 3 {7\over 2} {7\over 2} \right\rangle$ & $\left| 5 5 {9\over 2} {7\over 2} \right\rangle$ & $\left| 5 5 {11\over 2} {7\over 2} \right\rangle$ \\

&0.05 &  &  &  & {\bf 0.9657} & $-0.2555$ & $-0.0462$  \\
&0.22 &  &  &  & $-0.3213$ & {\bf 0.9382} & 0.1285 \\
&0.30 &  &  &  & $-0.2602$ &{\bf 0.9529} & 0.1560  \\

\noalign{\smallskip}\hline\noalign{\smallskip}

${9\over 2}[514]$ & & 
 &  &  &  & $\left| 5 5 {9\over 2} {9\over 2} \right\rangle$ & $\left| 5 5 {11\over 2} {9\over 2} \right\rangle$ \\

&0.05 &  &  &  & &{\bf 0.9657} & 0.0194 \\
&0.22 &  &  &  & &{\bf  0.9974} & 0.0716 \\
&0.30 &  &  &  & &{\bf 0.9959} & 0.0907 \\

\end{tabular}

\end{table*}

\begin{table*}[htb]

\caption{Correspondence between Nilsson orbitals $\Omega[N n_z \Lambda]$  and shell model orbitals $n l^j_\Omega$, as obtained by ``the rule'' of Eqs. (\ref{firstrule}) and (\ref{secondrule}). The quantum numbers $n_\rho$ and $\Sigma=\Omega-\Lambda$ in the cylindrical coordinates used by the Nilsson model and $n_r$ in the spherical coordinates used by the shell model are also shown. See Section 5 for further discussion. Below the name of each shell the orbitals belonging to it appear. }
\label{TC}       

\vskip 0.5cm 
\begin{tabular}{ r r r r r| r r r r r |r r r r r   }

$\Omega[N n_z \Lambda]$ & $n_\rho$ & $\Sigma$ & $nlj_\Omega$ & $n_r$ & 
$\Omega[N n_z \Lambda]$ & $n_\rho$ & $\Sigma$ & $nlj_\Omega$ & $n_r$ & 
$\Omega[N n_z \Lambda]$ & $n_\rho$ & $\Sigma$ & $nlj_\Omega$ & $n_r$ \\

\hline\noalign{\smallskip}
 sdg  & & & &  & pfh  & & & &  & sdgi & & & & \\ 
1/2[440] &0&+& 1g9/2$_{1/2}$ &0& 1/2[550] &0&+&1h11/2$_{1/2}$ &0& 1/2[660] &0&+&1i13/2$_{1/2}$ & 0\\
3/2[431] &0&+& 1g9/2$_{3/2}$ &0& 3/2[541] &0&+&1h11/2$_{3/2}$ &0& 3/2[651] &0&+&1i13/2$_{3/2}$ & 0\\
5/2[422] &0&+& 1g9/2$_{5/2}$ &0& 5/2[532] &0&+&1h11/2$_{5/2}$ &0& 5/2[642] &0&+&1i13/2$_{5/2}$ & 0\\
7/2[413] &0&+& 1g9/2$_{7/2}$ &0& 7/2[523] &0&+&1h11/2$_{7/2}$ &0& 7/2[633] &0&+&1i13/2$_{7/2}$ & 0\\
9/2[404] &0&+& 1g9/2$_{9/2}$ &0& 9/2[514] &0&+&1h11/2$_{9/2}$ &0& 9/2[624] &0&+&1i13/2$_{9/2}$ & 0\\
         & & &            & &11/2[505] &0&+&1h11/2$_{11/2}$ &0&11/2[615] &0&+&1i13/2$_{11/2}$ & 0\\
1/2[431] &0&$-$& 1g7/2$_{1/2}$ &0&          & & &          & &13/2[606] &0&+&1i13/2$_{13/2}$ & 0\\
3/2[422] &0&$-$& 1g7/2$_{3/2}$ &0& 1/2[541] &0&$-$& 1h9/2$_{1/2}$ &0&          & & &       &  \\
5/2[413] &0&$-$& 1g7/2$_{5/2}$ &0& 3/2[532] &0&$-$& 1h9/2$_{3/2}$ &0& 1/2[651] &0&$-$& 1i11/2$_{1/2}$ &0 \\
7/2[404] &0&$-$& 1g7/2$_{7/2}$ &0& 5/2[523] &0&$-$& 1h9/2$_{5/2}$ &0& 3/2[642] &0&$-$& 1i11/2$_{3/2}$ &0 \\
         & &   &            & & 7/2[514] &0&$-$& 1h9/2$_{7/2}$ &0& 5/2[633] &0&$-$& 1i11/2$_{5/2}$ &0 \\
1/2[420] &1&+& 2d5/2$_{1/2}$ &1& 9/2[505] &0&$-$& 1h9/2$_{9/2}$ &0& 7/2[624] &0&$-$& 1i11/2$_{7/2}$ &0 \\
3/2[411] &1&+& 2d5/2$_{3/2}$ &1&          & &   &            & & 9/2[615] &0&$-$& 1i11/2$_{9/2}$ &0 \\
5/2[402] &1&+& 2d5/2$_{5/2}$ &1& 1/2[530] &1&+& 2f7/2$_{1/2}$ &1&11/2[606] &0&$-$&1i11/2$_{11/2}$ &0 \\
         & & &            & & 3/2[521] &1&+& 2f7/2$_{3/2}$ &1&          & &   &            & \\
1/2[411] &1&$-$& 2d3/2$_{1/2}$ &1& 5/2[512] &1&+& 2f7/2$_{5/2}$ &1& 1/2[640] &1&+&2g9/2$_{1/2}$ &1 \\
3/2[402] &1&$-$& 2d3/2$_{3/2}$ &1& 7/2[503] &1&+& 2f7/2$_{7/2}$ &1& 3/2[631] &1&+&2g9/2$_{3/2}$ &1 \\
          & & &       & &          & & &       & & 5/2[622] &1&+&2g9/2$_{5/2}$ &1 \\ 
1/2[400] &2&+& 3s1/2$_{1/2}$ &2& 1/2[521] &1&$-$& 2f5/2$_{1/2}$ &1& 7/2[613] &1&+&2g9/2$_{7/2}$ &1 \\
          & & &       & & 3/2[512] &1&$-$& 2f5/2$_{3/2}$ &1& 9/2[604] &1&+&2g9/2$_{9/2}$ &1 \\
   s    & & &       & & 5/2[503] &1&$-$& 2f5/2$_{5/2}$ &1&          & & &       & \\
1/2[000] &0&+& 1s1/2$_{1/2}$ &0&           & & &       & & 1/2[631] &1&$-$& 2g7/2$_{1/2}$ &1 \\
          & & &       & & 1/2[510] &2&+& 3p3/2$_{1/2}$ &2& 3/2[622] &1&$-$& 2g7/2$_{3/2}$ &1 \\
 p         & & &       & & 3/2[501] &2&+& 3p3/2$_{3/2}$ &2& 5/2[613] &1&$-$& 2g7/2$_{5/2}$ &1 \\
 1/2[110] &0&+& 1p3/2$_{1/2}$ &0&               & & &       & & 7/2[604] &1&$-$& 2g7/2$_{7/2}$ &1 \\
 3/2[101] &0&+& 1p3/2$_{3/2}$ &0& 1/2[501] &2&$-$& 3p1/2$_{1/2}$ &2&          & & &            & \\
          & & &            & &          & &   &            & & 1/2[620] &2&+& 3d5/2$_{1/2}$ &2 \\
1/2[101] &0&$-$& 1p1/2$_{1/2}$ &0&  pf           & & &       & & 3/2[611] &2&+& 3d5/2$_{3/2}$ &2 \\
         & &  &             & &1/2[330] &0&+& 1f7/2$_{1/2}$ &0               & 5/2[602] &2&+& 3d5/2$_{5/2}$ &2 \\
 sd      & &  &             & &3/2[321] &0&+& 1f7/2$_{3/2}$ &0                &          & & &       & \\
1/2[220] &0& +& 1d5/2$_{1/2}$ &0&5/2[312] &0&+& 1f7/2$_{5/2}$ &0               & 1/2[611] &2&$-$& 3d3/2$_{1/2}$ &2 \\      
3/2[211] &0& +& 1d5/2$_{3/2}$ &0&7/2[303] &0&+& 1f7/2$_{7/2}$ &0  &     3/2[602] &2&$-$& 3d3/2$_{3/2}$ &2 \\
5/2[202] &0& +& 1d5/2$_{5/2}$ &0&                & & &       & &              & &   &            & \\
         & &  &            & &1/2[321] &0&$-$& 1f5/2$_{1/2}$ &0          & 1/2[600] &3&+& 4s1/2$_{1/2}$ &3 \\
1/2[211]&0&$-$& 1d3/2$_{1/2}$ &0&3/2[312] &0&$-$& 1f5/2$_{3/2}$ &0&  & & & & \\
3/2[202]&0&$-$& 1d3/2$_{3/2}$ &0&5/2[303] &0&$-$& 1f5/2$_{5/2}$ &0&   & & & & \\
         & & &             & &         & &   &            & &    & & & & \\ 
1/2[200] &1&+& 2s1/2$_{1/2}$  &1&1/2[310] &1&+& 2p3/2$_{1/2}$ &1& & & & & \\
         & & &             & &3/2[301] &1&+& 2p3/2$_{3/2}$ &1&    & & & & \\
         & & &             & &          & & &       & &         & & & & \\
         & & &             & &1/2[301] &1&$-$& 2p1/2$_{1/2}$ &1&  & & & & \\

\end{tabular}
\end{table*}

\begin{table*}[htb]

\caption{Matrix elements of the Nilsson Hamiltonian in the $N=2$ shell. All matrix elements possess a denominator equal to $\sqrt[6]{-(2 \delta +3)^2 (4 \delta -3)}$, shown here for brevity.  See Section 6 for further discussion.}
\label{TN2}       

\vskip 0.5cm 
\begin{tabular}{ r r r r r r r   }

N = 2 & $1/2[211-]$ & $1/2[200+]$ & $1/2[220+]$ & $3/2[202-]$ & $3/2[211+]$ & $5/2[202+]$ \\

\hline\noalign{\smallskip}

$1/2[211-]$ & $6.24 -0.58 \delta$  & $-0.26$ &  0.36 & 0. & 0. & 0. \\
$1/2[200+]$ & $-0.26$ & $1.15 (\delta +5.25)$ & 0. & 0. & 0. & 0. \\
$1/2[220+]$ & 0.36 & 0. & $-2.31 (\delta -2.63)$ & 0. & 0. & 0.\\
$3/2[202-]$ & 0. & 0. & 0. & $1.15 \delta +6.43$ & 0.36 & 0. \\
$3/2[211+]$ & 0. & 0. & 0. & 0.36 & $5.88 -0.58 \delta$  &0. \\
$5/2[202+]$ & 0. & 0. & 0. & 0. & 0. & $1.15 \delta +5.70$ \\

\end{tabular}
\end{table*}

\begin{table*}[htb]

\caption{Matrix elements of the Nilsson Hamiltonian in the $N=3$ shell. All matrix elements possess a denominator equal to $\sqrt[6]{-(2 \delta +3)^2 (4 \delta -3)}$, shown here for brevity.  See Section 6 for further discussion.}
\label{TN3}       
\vskip 0.5cm 
\begin{tabular}{ r r r r r r    }

N = 3 & $1/2[301-]$ & $1/2[321-]$ & $1/2[310+]$ & $1/2[330+]$ & $3/2[312-]$  \\ \hline

\hline\noalign{\smallskip}

$1/2[301-]$ & $\sqrt{3} (\delta +4.70)$ & 0.16 & 0.31 & 0. & 0. \\
$1/2[321-]$ & 0.16 & $7.91 -1.73 \delta$  & $-0.31$ & 0.38 & 0. \\
$1/2[310+]$ & 0.31 & $-0.31$ & 7.83 & 0.19 & 0. \\
$1/2[330+]$ & 0. & 0.38 & 0.19 & $7.91 -3.46 \delta$ & 0. \\
$3/2[312-]$ & 0. & 0. & 0. & 0. & 7.99  \\
$3/2[301+]$ & 0. & 0. & 0. & 0. & $-0.22$  \\
$3/2[321+]$ & 0. & 0. & 0. & 0. & 0.44  \\
$5/2[303-]$ & 0. & 0. & 0. & 0. & 0. \\
$5/2[312+]$ & 0. & 0. & 0. & 0. & 0. \\
$7/2[303+]$ & 0. & 0. & 0. & 0. & 0.\\

\end{tabular}
\vskip 0.5cm 

\begin{tabular}{ r r r r r r   }

N = 3 & $3/2[301+]$ & $3/2[321+]$ & $5/2[303-]$ & $5/2[312+]$ & $7/2[303+]$ \\ \hline

\hline\noalign{\smallskip}

$1/2[301-]$ & 0. & 0. & 0. & 0. & 0. \\
$1/2[321-]$ & 0. & 0. & 0. & 0. & 0. \\
$1/2[310+]$ & 0. & 0. & 0. & 0. & 0. \\
$1/2[330+]$ & 0. & 0. & 0. & 0. & 0. \\
$3/2[312-]$ & $-0.22$ & 0.44 & 0. & 0. & 0. \\
$3/2[301+]$ & $1.73 \delta+7.83$ & 0.16 & 0. & 0. & 0. \\
$3/2[321+]$ & 0.16 & $7.60 -1.73 \delta$ & 0. & 0. & 0. \\
$5/2[303-]$ & 0. & 0. & $\sqrt{3}(\delta +4.70)$& 0.38 & 0. \\
$5/2[312+]$ & 0. & 0. & 0.38 & 7.37 & 0. \\
$7/2[303+]$ & 0. & 0. & 0. & 0. & $1.73 \delta +7.21$ \\

\end{tabular}
\end{table*}

\end{document}